\documentclass[10pt,twocolumn]{article}

\usepackage{graphicx,amsmath,amsfonts,amssymb,dcolumn,amsthm,slashed,bbm,xspace,pgffor,tikz,mathbbol,latexsym,enumerate}
\usepackage{soul}
\usepackage[colorlinks,citecolor=red,urlcolor=cyan,linkcolor=blue]{hyperref}
\usepackage[utf8]{inputenc}
\usepackage[english]{babel}
\usepackage{lmodern}
\usepackage{microtype}
\usepackage{cite}

\numberwithin{equation}{section}

\begin{document}

\title{\textbf{Master equations governing the coupling between spin-currents and gravity}}

\author{\textbf{A.~de Camargo}$^a$\thanks{	alexandrec@id.uff.br}, \  \textbf{R.~F.~Sobreiro}$^a$\thanks{rodrigo\_sobreiro@id.uff.br} \ and \textbf{V.~J.~Vasquez Otoya}$^b$\thanks{victor.vasquez@ifsudestemg.edu.br}\\\\
\textit{$^a${\small UFF - Universidade Federal Fluminense, Instituto de F\'isica,}}\\
\textit{{\small Av. Litorânea s/n, 24210-346, Niter\'oi, RJ, Brasil.}}\\
\textit{$^b${\small IFSEMG - Instituto Federal de Educa\c c\~ao, Ci\^encia e Tecnologia do Sudeste de Minas Gerais,}}\\
\textit{{\small Rua Bernardo Mascarenhas 1283, 36080-001,
Juiz de Fora, MG, Brasil.}}}

\date{}
\maketitle

\begin{abstract}
In this work we consider the Einstein-Hilbert action in the first order order formalism coupled to Dirac spinors. From the little group symmetry, we derive the corresponding Bargmann-Wigner current which is conserved but not gauge invariant. Therefore, we construct a gauge invariant version of the Bargmann-Wigner current which is not conserved but potentially observable. Because it is not conserved we split this current into fermionic and gravitational sectors and derive their broken continuity equations for each sector. These equations compose the master equations governing the interaction of spin-currents with gravity. Furthermore, we derive the corresponding master equations in the weak field approximation.
\end{abstract}

\section{Introduction}

In condensed matter, spin currents are governed by a broken continuity equation \cite{Vernes:2007prb} with many possible applications in technology \cite{Awschalom:2002sca,Fert:2008zz}. A relativistic description of spin-currents in the context of Dirac fermions coupled to the electromagnetic fields was developed in \cite{Vernes:2007prb} where the authors derive a relativistic broken continuity equation for spin-currents. The usual broken equation is then obtained from its non-relativistic limit at leading order. The authors also show that the next-to-leading order corrections contributes to spin-transfer torque effects. See also \cite{Sokolov:1986}.

Following \cite{Vernes:2007prb}, the authors of \cite{Sobreiro:2011ie} provided a gauge theoretical point of view of the spin-current relativistic broken equation. Specifically, they show that Noether current for the little group\footnote{The little group is a subgroup of the Poincaré group which characterizes the spin of the fields \cite{Bargmann:1948ck}. See Section \ref{SYM}.}, known as the Bargmann-Wigner current \cite{Bargmann:1948ck}, is not gauge invariant. Thus, due to the gauge principle \cite{tHooft:1994pio,Rubakov:2002fi}, such current cannot be an observable. Hence, by evoking the gauge principle and the minimal coupling, an observable spin-current is derived and the broken continuity equation of \cite{Vernes:2007prb} is recovered. In \cite{Santos:2012ar}, this result was generalized for Dirac-Maxwell electrodynamics in material media. See also \cite{Wang:2006vhp} for a discussion about spin currents in the context of Noether currents.

In the present work we consider Dirac fermions coupled with the gravitational field in the first order formalism \cite{Utiyama:1956sy,Kibble:1961ba,Sciama:1964wt,DeSabbata:1986sv,Mardones:1990qc,Zanelli:2005sa} and derive the master equations governing the coupling of spin-currents with the gravitational field. Starting with the Dirac-Einstein-Hilbert action with no cosmological constant and following the prescription of \cite{Sobreiro:2011ie}, a broken continuity equation for spin-currents is derived. Due to the different nature of the Einstein-Hilbert action in comparison with the Maxwell one, the coupling obtained carry some different components from the the way spin-currents couple to electromagnetic fields. The complete setting for looking for gravitational effects over spin-currents is then provided.

The work is organized as follows: In Section \ref{EHD}, we present the fields, the action and the field equations. In Section \ref{SYM}, we derive the symmetries and currents relevant to this work. Section \ref{MASTER} is dedicated to obtain the master equations ruling spin-current-gravity interaction. In Section \ref{WEAK}, we particularize the master equations to the weak gravitational field case. Finally, our conclusions are displayed in Section \ref{FINAL}.

\section{Einstein-Cartan gravity coupled to Dirac fermions}\label{EHD}

We will consider in this section general relativity coupled to fermions in the first order formalism. In such description, the gravity sector is formulated in terms of the vierbein and spin-connection and the gravity action we employ is simply the Einstein-Hilbert one with no cosmological constant, see \cite{Utiyama:1956sy,Kibble:1961ba,Sciama:1964wt,Zanelli:2005sa}. The fermionic sector will be described by Dirac fermions whose dynamics is governed by the Dirac action minimally coupled to gravity \cite{DeSabbata:1986sv,Jacobson:1988qta,Hehl:1994ue}.

We start by considering a generic 4-dimensional Riemmann-Cartan differential manifold $M$ describing spacetime. A coordinate system in a point $x\in M$ is labeled by $x^\mu$ (Greek indices refer to world indices and run through $\{0,1,2,3\}$). The equivalence principle is realized by imposing that at each point $x\in M$, an inertial coordinate system $x^a$ (Lower case Latin indices refer to frame indices and also run through $\{0,1,2,3\}$) exists. The vierbein field $e^a_\mu$ and its inverse $e_a^\mu$ ensures the validity of the equivalence principle for all points in $M$ by means of
\begin{eqnarray}
dx^a&=&e^a_\mu dx^\mu\;,\nonumber\\
dx^\mu&=&e^\mu_adx^a\;.\label{e1}
\end{eqnarray}
Obviously, there is an infinity number of possible inertial frames, all related to each other by a Lorentz transformation. For instance, given two frames ${x^\prime}^a$ and $x^a$, they are related to each other through a local Lorentz transformation\footnote{Rigorously, all inertial frames are related through Poincaré transformations. Nevertheless, since the Poincaré group is not semi-simple, we construct only a gauge theory for the Lorentz group. The translational sector, however, is not discarded because it can be interpreted as a vector representation of the Lorentz group.} $\Lambda(x)\in SO(1,3)$ by means of ${dx^\prime}^a=\Lambda^a_{\phantom{a}b}dx^b$. Since this transformation does not affect coordinates $x^\mu$, it is a genuine gauge transformation. Moreover, the Lorentz group is the isometry group of a Minkowski space. In practice, one typically identifies the tangent space $T_x(M)$ at each point $x\in M$ with the space of all inertial frames at the very same point. Hence, the vierbein defines a local isomorphism $M\longmapsto T(M)\sim\mathbb{R}^{(1,3)}$. In $T_x(M)$, the Minkowski metric $\eta_{ab}$ and its inverse $\eta^{ab}$ are defined here with positive signature. Relations \eqref{e1} automatically imply on the existence of the metric field and its inverse
\begin{eqnarray}
    g_{\mu\nu}&=&\eta_{ab}e^a_\mu e^b_\nu\;,\nonumber\\
    g^{\mu\nu}&=&\eta^{ab}e^\mu_ae^\nu_b\;.\label{g1}
\end{eqnarray}
These relations are easily obtained from the invariance of the line element $ds^2=\eta_{ab}dx^adx^b=g_{\mu\nu}dx^\mu dx^\nu$. Moreover, the following inverse relations are always true,
\begin{eqnarray}
    e^a_\mu e^\mu_b&=&\delta^a_b\;,\nonumber\\
    e^a_\mu e^\nu_a&=&\delta^\nu_\mu\;.\label{inverse1}
\end{eqnarray}
Thence, the vierbein field encodes all metric properties of $M$.

A second mathematical object can be defined, the spin-connection $\omega^a_{\mu b}$, characterizing the parallelism properties of $M$ between near points. In fact, just like the vierbein is related to the metric tensor, the spin-connection is related to the affine connection through
\begin{equation}
    \Gamma^\mu_{\alpha\beta}=e_a^\mu\left(\partial_\alpha e_\beta^a+\omega_{\alpha b}^ae^b_\beta\right)\;.\label{affine1}
\end{equation}
This relation is also known as the compatibility condition, ensuring the vanishing of non-metricity. In the context of a gauge theory for the local Lorentz group the spin-connection emerges naturally as the gauge field of the theory. Hence, the double indices in the tangent space is associated with the fact that the spin-connection is an algebra-valued field in the adjoint representation of the Lorentz group. As a consequence, $\omega_{\mu}^{ab}=\omega_{\mu}^{ba}$.

In the gauge approach we are constructing, the vierbein and the spin-connection are regarded as the fundamental gravitational fields. The infinitesimal local Lorentz transformations of these fields are defined by
\begin{eqnarray}
    \delta\omega^{ab}_\mu&=&D_\mu\alpha^{ab}\;,\nonumber\\
\delta e^a_\mu&=&\alpha^a_{\phantom{a}b}e^b_\mu\;,\nonumber\\
\delta e^\mu_a&=&-\alpha_a^{\phantom{a}b}e^\mu_b\;,\label{gauget1}
\end{eqnarray}
with $\alpha(x)$ being an infinitesimal local parameter and $D_\mu\cdot^a\equiv\partial_\mu\cdot^a+\omega^a_{\mu b}\cdot^b$ is the covariant derivative in the adjoint representation. It is clear from \eqref{gauget1} that the fundamental fields are not gauge invariant. Thence, from the gauge principle, they can not be observables. The observables of the theory must be gauge invariant objects. In fact, one can easily check that the metrics \eqref{g1} and affine connection \eqref{affine1} are examples of gauge invariant objects. Equipped with $e^a_\mu$ and $\omega^a_{\mu b}$, one can define the field strengths of the theory, namely, the curvature and torsion tensors, respectively given by,
\begin{eqnarray}
    R^{ab}_{\mu\nu}&=&\partial_\mu\omega^{ab}_\nu-\partial_\nu\omega^{ab}_\mu+\omega^a_{\phantom{a}c\mu}\omega^{cb}_\nu-\omega^a_{\phantom{a}c\nu}\omega^{cb}_\mu\;,\nonumber\\
    T^a_{\mu\nu}&=&D_\mu e_\nu^a-D_\nu e_\mu^a\;,\label{fieldst1}
\end{eqnarray}
which transform covariantly under gauge transformations.

At this point we introduce fermions in the theory. Precisely, Dirac spinors representing, for instance, electrons. The Clifford algebra can be defined in tangent space, $\{\gamma^a,\gamma^b\}=2\eta^{ab}$, with $\gamma^a$ being the gamma matrices in Dirac representation \cite{Itzykson:1980rh}. The fermions are represented by the spinor field $\psi$ and its conjugate $\bar{\psi}=\psi^\dagger\gamma^0$. Under infinitesimal gauge transformations, these fields transform by
\begin{eqnarray}
\delta\psi&=&-i\alpha^{ab}\sigma_{ab}\psi\;,\nonumber\\
\delta\bar{\psi}&=&i\bar{\psi}\alpha^{ab}\sigma_{ab}\;,\label{gauget2}
\end{eqnarray}
with $\sigma_{ab}=\dfrac{i}{2}[\gamma_a,\gamma_b]$. The fermionic fields constitute the fundamental matter sector of the model.

With all ingredients defined, we can now define the theory governing the dynamics of all fields. Gravity coupled with Dirac spinors is then described by the Einstein-Hilbert-Dirac action,
\begin{eqnarray}
    S&=&-\frac{1}{2k}\int d^4x\;\mathrm{e}R^{ab}_{\mu\nu}e^\mu_ae^\nu_b+\nonumber\\
    &+&\int d^4x\;\mathrm{e}\bar{\psi}\left(i\gamma^ae^\mu_a D_\mu-m\right)\psi\;,\label{action1}
\end{eqnarray}
where $k=1/8\pi G$ with $G$ being Newton's constant and $m$ standing for the fermionic mass. The covariant derivative in the fundamental representation reads $D_\mu=\partial_\mu-\dfrac{i}{4}\sigma_{ab}\omega^{ab}_\mu$. Needless to say, action \eqref{action1} is invariant under gauge transformations \eqref{gauget1} and \eqref{gauget2}.

The field equations can be easily computed from \eqref{action1}. For the vierbein field we get the well known Einstein equation in the first order formalism
\begin{equation}
R^a_\mu-\frac{1}{2}Re^a_\mu=ik\bar{\psi}\gamma^aD_\mu\psi\;,\label{feq1}
\end{equation}
with
\begin{eqnarray}
    R^a_\mu&=&R^{ab}_{\mu\nu}e^\nu_b\;,\nonumber\\
    R&=&R^a_\mu e^\mu_a\;,\label{RicciR}
\end{eqnarray}
being respectively the Ricci tensor and the curvature scalar. For the spin-connection we get,
\begin{equation}
    D_\nu\left[\mathrm{e}\left(e_a^\mu e_b^\nu-e_b^\mu e_a^\nu\right)\right]=k\mathrm{e}\bar{\psi}\gamma^ce^\mu_c\sigma_{ab}\psi\;.\label{feq2}
\end{equation}
For the $\bar{\psi}$, one easily gets the usual Dirac equation in a curved spacetime
\begin{equation}
\left(i\gamma^ae_a^\mu D_\mu-m\right)\psi=0\;.\label{psibar}
\end{equation}
From \eqref{psibar} one can find
\begin{equation}
\bar{\psi}\left(i\tilde{D}_\mu\gamma^ae_a^\mu+m\right)=0\;,\label{psi1}
\end{equation}
with $\tilde{D}_\mu=\overset{\leftarrow}{\partial}_\mu+\dfrac{i}{4}\sigma_{ab}\omega^{ab}_\mu$. 
On the other hand, the field equation for $\psi$ computed directly from the EHD action \eqref{action1} is given by
\begin{equation}
i\left(\mathrm{e}\bar{\psi}\gamma^ae_a^\mu\right)\tilde{D}_\mu+\mathrm{e}m\bar{\psi}=0\;.\label{psi2}
\end{equation}
Combination of \eqref{psi1} and \eqref{psi2} gives the interesting relation
\begin{equation}
    D_\mu\left(\mathrm{e}e_a^\mu\right)=0\;.\label{e}
\end{equation}
Combination of equation \eqref{feq1} with the second equation of \eqref{psibar}, gives the curvature scalar in terms of the matter content,
\begin{equation}
    R=-mk\bar{\psi}\psi\;.\label{R1}
\end{equation}

\section{Symmetries and currents}\label{SYM}

In this section we summarize the main currents relevant to this work, namely, gauge current, chiral currents, and the Bargmann-Wigner current. The most relevant symmetry of a gauge theory is, as expected, the gauge symmetry whose transformations are given in \eqref{gauget1} and \eqref{gauget2}.
The corresponding Noether current reads
\begin{equation}
    j_{ab}^\mu=\mathrm{e}\bar{\psi}\gamma^ce_c^\mu\sigma_{ab}\psi\;,\label{c1}
\end{equation}
which is recognized as the spin density appearing in equation \eqref{feq2}.

For massless fermions, chiral symmetry is also present. This is not the case of action \eqref{action1}, but chiral currents will emerge at some point in our computations. In fact, we have two types of chiral currents to discuss. The first chiral transformations are
\begin{eqnarray}
\delta_c^{(1)}\psi&=&-i\alpha\gamma^5\psi\;,\nonumber\\
\delta_c^{(1)}\bar{\psi}&=&-i\alpha\bar{\psi}\gamma^5\;,\label{chi1}
\end{eqnarray}
and zero for all other fields. The scalar $\alpha$ is a global parameter. The corresponding non-conserved current is the chiral current characterized by
\begin{eqnarray}
S^\mu&=&\mathrm{e}e^\mu_a\bar{\psi}\gamma^a\gamma^5\psi\;=\;\mathrm{e}e^\mu_aS^a\;,\nonumber\\
\partial_\mu S^\mu&=&2im\mathrm{e}\bar{\psi}\gamma^5\psi\;.\label{chi2}
\end{eqnarray}
Clearly, the second of equations \eqref{chi2} states the non-conservation of the first chiral current due to the non-vanishing of the fermion mass.

A second chiral transformation can be defined as well,
\begin{eqnarray}
\delta_c^{(2)}\psi&=&-i\alpha^{ab}\gamma^5\sigma_{ab}\psi\;,\nonumber\\
\delta_c^{(2)}\bar{\psi}&=&-i\alpha^{ab}\bar{\psi}\gamma^5\sigma_{ab}\;,\label{chi3}
\end{eqnarray}
providing,
\begin{eqnarray}
S_\mu^{ab}&=&\mathrm{e}e_\mu^c\bar{\psi}\gamma_c\gamma^5\sigma^{ab}\psi\;,\nonumber\\
D^\mu S_\mu^{ab}&=&2im\mathrm{e}\bar{\psi}\gamma^5\sigma^{ab}\psi\;,\label{chi4}
\end{eqnarray}
The non-conservation of the second chiral current is also evident from the second equation in \eqref{chi4}.

We now proceed to discuss a symmetry of great relevance in this paper, the little group symmetry \cite{Bargmann:1948ck,Itzykson:1980rh,Sobreiro:2011ie,Santos:2012ar,Kryuchkov:2015acw}. Let us consider the following local subgroup sequence associated with spacetime automorphisms of the fields:
\begin{equation}
    \mathrm{diff}(1,3)\longrightarrow A(1,3;\mathbb{R})\longrightarrow ISO(1,3)\;,\label{seq1}
\end{equation}
determining the local Poincaré group as a stability group of the diffeomorphism group. In \eqref{seq1}, $A(1,3;\mathbb{R})=GL(1,3;\mathbb{R})\times\mathbb{R}^{1,3}$ is the affine group. Thence, we consider a second subgroup sequence:
\begin{equation}
    ISO_{local}(1,3)\longrightarrow ISO_{global}(1,3)\longrightarrow L(1,3)\;.
\end{equation}
where $L(1,3)$ stands for the little group, a stability group of the Poincaré group \cite{Bargmann:1948ck}. Therefore, the very restricted class of global little group transformations, which defines spin and spin currents (see below), is contained in the full diffeomorphism group. Consequently, we will be able to define spin and spin currents in the same way we define them in Minkowski spacetime. Physically, the little group is a particular Lorentz transformation which preserves linear momentum. The corresponding generator is the Pauli-Lubanski pseudo-vector \cite{Bargmann:1948ck,Sobreiro:2011ie},
\begin{equation}
    W^\mu=-\frac{1}{2}\epsilon^{\mu\nu\alpha\beta}J_{\nu\alpha}P_\beta\;,\label{pl1}
\end{equation}
with $J_{\mu\nu}$ being the internal angular momentum (spin) of the corresponding field and $P_\mu=i\partial_\mu$ being the generators of translations. 
Because the spin differs from field to field, we need to specify $J_{\mu\nu}$ for each field. For spinors, $J_{\mu\nu}=e^a_\mu e^b_\nu\sigma_{ab}/2$, implying that the fermionic Pauli-Lubanski pseudo-vector is given by
\begin{equation}
    W^\mu_F=-\frac{1}{2}\epsilon^{\mu\nu\alpha\beta}e_\nu^ae_\alpha^b\sigma_{ab}P_\beta\;.\label{pl2}
\end{equation}
For vector fields, we can use $\left(J_{\mu\nu}\right)_{\alpha\beta}=(\eta_{\mu\alpha}\eta_{\nu\beta}-\eta_{\mu\beta}\eta_{\nu\alpha})/2$. The corresponding Pauli-Lubanski pseudo-tensor reads
\begin{equation}
    W^{\mu\nu\alpha}_B=-\frac{1}{2}\epsilon^{\mu\nu\alpha\beta}P_\beta\;.\label{pl3}
\end{equation}

The little group transformations for the dynamical fields in action \eqref{action1} are given by\footnote{The identity $\gamma^5\sigma_{\mu\nu}=\frac{i}{2}\epsilon_{\mu\nu\alpha\beta}\sigma^{\alpha\beta}$, with $\sigma_{\mu\nu}=e_\mu^ae_\nu^b\sigma_{ab}$, was employed.}
\begin{eqnarray}
\delta_l\omega^{ab}_\mu&=&-\frac{i}{2}\xi_\nu\epsilon_\mu^{\phantom{\mu}\nu\alpha\beta}P_\beta\omega_\alpha^{ab}\;,\nonumber\\
\delta_l e^a_\mu&=&-\frac{i}{2}\xi_\nu\epsilon_\mu^{\phantom{\mu}\nu\alpha\beta}P_\beta h_\alpha^a\;,\nonumber\\
\delta_l e^\mu_a&=&-\frac{i}{2}\xi_\nu\epsilon^{\mu\nu\phantom{\alpha}\beta}_{\phantom{\mu\nu}\alpha}P_\beta H^\alpha_a\;,\nonumber\\
\delta_l\psi&=&\frac{1}{2}\xi_\mu\gamma^5\sigma^{ab}e^\mu_ae_b^\nu P_\nu\psi\;,\nonumber\\
\delta_l\bar{\psi}&=&\frac{1}{2}\bar{\psi}\overset{\leftarrow}{P}_\nu\xi_\mu\gamma^5\sigma^{ab}e^\mu_ae_b^\nu\;,\label{stab3}
\end{eqnarray}
where $\xi_\mu$ is a global parameter. The corresponding Noether current, i.e., the Bargmann-Wigner current, is given by
\begin{eqnarray}
    K^{\mu\nu}&=&\frac{\mathrm{e}}{4k}e^\mu_ae^\alpha_b\epsilon_\alpha^{\phantom{\alpha}\nu\beta\gamma}\left(\partial_\beta\omega^{ab}_\gamma-\partial_\gamma\omega^{ab}_\beta\right)+\nonumber\\
    &-&\frac{\mathrm{e}}{2}e^\mu_ce^\nu_ae^\alpha_b\bar{\psi}\gamma^c\gamma^5\sigma^{ab}\partial_\alpha\psi\;.\label{bw1}
\end{eqnarray}
A remarkable feature of this current is due to the fact that in the action \eqref{action1} there are no derivatives of the vierbein fields whatsoever. Thus, there is no contribution from the vierbein in the Bargmann-Wigner current \eqref{bw1}. In fact, the first term in \eqref{bw1} is the contribution coming from the spin connection and the second one is the contribution coming from the electron. The Bargmann-Wigner current is clearly conserved, $\partial_\mu T^{\mu\nu}=0$, but not locally gauge invariant. Hence, due to the gauge principle, it cannot be an observable of the theory.

\section{Gauge principle and spin current master equations}\label{MASTER}

Following \cite{Vernes:2007prb,Sobreiro:2011ie,Santos:2012ar} the idea is to construct, by induction, a gauge invariant version of the Bargmann-Wigner current \eqref{bw1}. The result should be an observable of the model, as established by the gauge principle \cite{Itzykson:1980rh,Rubakov:2002fi}, but will not be conserved any more. In the case of electrodynamics coupled with fermions \cite{Vernes:2007prb,Sobreiro:2011ie,Santos:2012ar}, this was done by imposing the minimal coupling rule $\partial\longrightarrow D$. In the present case we need extra careful because we are dealing with a non-Abelian gauge theory with non-vanishing background. Henceforth, our \emph{ansatz} is to perform the following substitutions in \eqref{bw1}
\begin{eqnarray}
    \partial_\alpha\psi&\longrightarrow&D_\alpha\psi\;,\nonumber\\
    \partial_\beta\omega^{ab}_\gamma-\partial_\gamma\omega^{ab}_\beta&\longrightarrow&R^{ab}_{\beta\gamma}\;.\label{mincoup1}
\end{eqnarray}
The first one is the minimal coupling rule employed in \cite{Sobreiro:2011ie,Santos:2012ar}. The second one is intuitively the natural choice. The reason is that in the Abelian example of electrodynamics (or in any Abelian gauge theory), the quantity $ \partial_\beta\omega^{ab}_\gamma-\partial_\gamma\omega^{ab}_\beta$ is indeed the field strength. Hence, the gauge invariant current we are looking for is given by
\begin{equation}
    \mathcal{K}^{\mu\nu}=\frac{\mathrm{e}}{4k}e^\mu_ae^\alpha_b\epsilon_\alpha^{\phantom{\alpha}\nu\beta\gamma}R^{ab}_{\beta\gamma}-\frac{\mathrm{e}}{2}e^\mu_ce^\nu_ae^\alpha_b\bar{\psi}\gamma^c\gamma^5\sigma^{ab}D_\alpha\psi\;.\label{sc1}
\end{equation}
Since this current is not conserved anymore, it can be composed by two distinguished gauge invariant parts, the bosonic and the fermionic one, respectively written as
\begin{eqnarray}
    B^{\mu\nu}&=&\frac{\mathrm{e}}{2k}e^\mu_a\tilde{R}^{a\nu}\;,\nonumber\\
    F^{\mu\nu}&=&\frac{\mathrm{e}}{2}e^\mu_ce^\nu_ae^\alpha_b\bar{\psi}\gamma^c\gamma^5\sigma^{ab}D_\alpha\psi\;,\label{sc2}
\end{eqnarray}
with
\begin{eqnarray}
    \tilde{R}^{ab}_{\mu\nu}&=&\frac{\mathrm{e}}{2}\epsilon_{\mu\nu}^{\phantom{\mu\nu}\alpha\beta}R^{ab}_{\alpha\beta}\;,\nonumber\\
    \tilde{R}^{a\mu}&=&e_b^\alpha\tilde{R}^{ab\;\mu}_\alpha\;.
\end{eqnarray}
We remark that, for $e^\mu_a\equiv\eta^\mu_a$, characterizing a fixed Minkowski spacetime, the fermionic current in \eqref{sc2} reduces, in form, to the one in \cite{Vernes:2007prb,Sobreiro:2011ie}. Nevertheless, due to the fact that the EH action differs from the YM one, the bosonic sector is actually different from that of \cite{Sobreiro:2011ie}. This situation would correspond to a gauge theory for the Lorentz group with the absence of the vierbein, \emph{i.e.,} a theory where the equivalence principle does not apply.

The on-shell divergence of the currents \eqref{sc2} can be straightforwardly computed. For the bosonic sector, and with the help of equation \eqref{feq2}, one readily gets
\begin{equation}
    \partial_\mu B_\nu^\mu=\frac{1}{4}j^\alpha_{ab}\tilde{R}^{ab}_{\alpha\nu}+\frac{\mathrm{e}}{4k}\left(e^\mu_ae^\alpha_b-e^\alpha_ae^\mu_b\right)D_\mu\tilde{R}^{ab}_{\alpha\nu}\;.\label{master1}
\end{equation}
The fermionic sector demands a bit more work. First, we rewrite it as
\begin{equation}
    F^{\mu\nu}=-\frac{m}{2}\mathrm{e}e^\mu_ae^\nu_b\bar{\psi}\gamma^a\gamma^5\gamma^b\psi+\frac{i}{2}\mathrm{e}\bar{\psi}e^\mu_c\gamma^c\gamma^5D^\nu\psi\;,\label{spinf2}
\end{equation}
where the field equation \eqref{feq2} was employed. Then,
\begin{eqnarray}
    \partial_\mu F^{\mu\nu}&=&\frac{i}{8}R^{\mu\nu}_{ab}S^{ab}_\mu+\frac{m}{2}\mathrm{e}e^\mu_aD_\mu e^\nu_b\bar{\psi}\gamma^5\gamma^a\gamma^b\psi+\nonumber\\
    &+&\frac{i}{2}\mathrm{e}D^\nu e^\mu_a\bar{\psi}\gamma^5\gamma^aD_\mu\psi\;.\label{master2a}
\end{eqnarray}   
Employing \eqref{e} and the identity
\begin{equation}
    D_\mu\mathrm{e}=-\mathrm{e}e^b_\nu D_\mu e^\nu_b\;,
\end{equation}
we finally get
\begin{equation}
    \partial_\mu F^{\mu\nu}=\frac{i}{8}R^{\mu\nu}_{ab}S^{ab}_\mu-\frac{i}{4}\Theta^{ab}_5T_{ab}^\nu+X^\nu\;,\label{master2b}
\end{equation} 
with
\begin{eqnarray}
    T^\mu_{ab}&=&e^\nu_aD_\nu e^\mu_b-e^\nu_bD_\nu e^\mu_a\;,\nonumber\\
    \Theta^{ab}_5&=&m\mathrm{e}\bar{\psi}\gamma^5\sigma^{ab}\psi\;,\nonumber\\
    X^\nu&=&\frac{i}{2}D^\nu\left(\mathrm{e}e^\mu_a\right)\bar{\psi}\gamma^5\gamma^aD_\mu\psi\label{tensors1}
\end{eqnarray}
The bilinear $\Theta^{ab}_5$ is an antisymmetric pseudo-tensor which also appears in \eqref{chi4}. Moreover, the quantity $T^{ab}_\mu$ is related to the torsion 2-form through
\begin{equation}
    T^\mu_{ab}=-e^\alpha_ae^\beta_be^\mu_cT^c_{\alpha\beta}\;.
\end{equation}

It is important to notice that the derivation of equation \eqref{master2b} uses only the field equations of the spinors, namely \eqref{psibar}, \eqref{psi1}, and \eqref{psi2}. Thence, equation \eqref{master2b} is indeed valid for external gravitational fields as well. In other words, it also describes the situation of free spin-currents immersed in a background gravitational field. Moreover, the interaction between electrons due to the electromagnetic field is not considered. Hence, equation \eqref{master2b} is suitable for one electron systems or weakly interacting systems.

\section{Weak field approximation}\label{WEAK}

The weak field approximation is taken from the usual linearization of the vierbein:
\begin{eqnarray}
    e^a_\mu&\approx&\eta^a_\mu+h^a_\mu\;,\nonumber\\
    e_a^\mu&\approx&\eta_a^\mu-h_a^\mu\;,\label{linear1}
\end{eqnarray}
with $\{\eta^a_\mu,\eta_a^\mu\}$ being the Minkowski vierbeins defining the spacetime background to be the flat Minkowski manifold. The fields $\{h^a_\mu,h_a^\mu\}$ are the gravitational small fluctuations that we can refer to as gravitons. The vierbein determinant simplifies to $\mathrm{e}\approx1+\mathrm{h}$ with $\mathrm{h}=\eta^\mu_ah^a_\mu$. The spin-connection is also perturbed around a Minkowski spacetime. Thence, we can write
\begin{equation}
    \omega=\underline{\omega}(\eta+h)+\bar{\omega}\approx\frac{\partial\underline{\omega}}{\partial h^a_\mu}\bigg|_{h=0}h^a_\mu+\bar{\omega}=z_a^\mu h^a_\mu+\bar{\omega}\;,\label{omega1}
\end{equation}
with $\bar{\omega}$ being the metric independent part of the spin-connection and $\underline{\omega}(\eta)=0$. Moreover, $\bar{\omega}$ is taken as a small fluctuation. From \eqref{affine1}, the linearization of the affine connection can be found,
\begin{equation}
   \Gamma^\mu_{\alpha\beta}\approx\eta_a^\mu\left(\partial_\alpha h_\beta^a+z^{a\sigma}_{\alpha bc}h^c_\sigma\eta^b_\beta\right)+\eta^\mu_a\bar{\omega}_{\alpha b}^a\eta^b_\beta\;,\label{affine2}
\end{equation}
with $z^{a\sigma}_{\alpha bc}=\frac{\partial\underline{\omega}^a_{\alpha b}}{\partial h^c_\sigma}\bigg|_{h=0}$ being a constant tensorial structure. Using \eqref{omega1}, we can find the linear curvature and its dual,
\begin{eqnarray}
    R^{ab}_{\mu\nu}(\omega)&\approx&\partial_\mu\bar{\omega}^{ab}_\nu-\partial_\nu\bar{\omega}^{ab}_\mu+\nonumber\\
    &+&z^{ab\sigma}_{\nu c}\partial_\mu h^c_\sigma-z^{ab\sigma}_{\mu c}\partial_\nu h^c_\sigma\;,\nonumber\\
    \tilde{R}^{ab}_{\mu\nu}(\omega)&\approx&\epsilon_{\mu\nu}^{\phantom{\mu\nu}\alpha\beta}\partial_\alpha\bar{\omega}^{ab}_\beta+\epsilon_{\mu\nu}^{\phantom{\mu\nu}\alpha\beta}z^{ab\sigma}_{\beta c}\partial_\alpha h^c_\sigma\;.\nonumber\\
    \label{lincurv1}
\end{eqnarray}
The linearized dual Ricci tensor then reads
\begin{equation}
    \tilde{R}^a_\nu\approx\eta^\mu_b\epsilon_{\mu\nu}^{\phantom{\mu\nu}\alpha\beta}\partial_\alpha\bar{\omega}^{ab}_\beta+\eta^\mu_b\epsilon_{\mu\nu}^{\phantom{\mu\nu}\alpha\beta}z^{ab\sigma}_{\beta c}\partial_\alpha h^c_\sigma\;.\label{linricci1}
\end{equation}
 
Let us start with the bosonic master equation \eqref{master1}. First, we consider the bosonic gauge invariant Bargmann-Wigner current
\begin{equation}
    B^{\mu\nu}\approx\frac{1}{2k}\eta^\mu_a\eta^\gamma_b\left[\epsilon_\gamma^{\phantom{\gamma}\nu\alpha\beta}\partial_\alpha\bar{\omega}^{ab}_\beta+\epsilon_\gamma^{\phantom{\gamma}\nu\alpha\beta}z^{ab\sigma}_{\beta c}\partial_\alpha h^c_\sigma\right]\;,\label{bbwlin1}
\end{equation}
In the same spirit, the gauge current \eqref{c1} reads
\begin{equation}
    j_{ab}^\mu\approx G^\mu_{ab}+\mathrm{h}G^\mu_{ab}-h^\mu_cG^c_{ab}\;,\label{c1a}
\end{equation}
with $G^c_{ab}=\bar{\psi}\gamma^c\sigma_{ab}\psi$ and $G^\mu_{ab}=\eta^\mu_cG^c_{ab}$. Considering only linear terms in this approximation, equation \eqref{master1} reduce to
\begin{eqnarray}
    \partial_\mu B_\nu^\mu&=&\frac{1}{4}\epsilon_{\alpha\nu}^{\phantom{\alpha\nu}\beta\gamma}G^\alpha_{ab}\partial_\beta\bar{\omega}^{ab}_\gamma+\nonumber\\
    &+&\frac{\mathrm{1}}{2k}\epsilon_{\alpha\nu}^{\phantom{\alpha\nu}\beta\gamma}\eta^\mu_a\eta^\alpha_b\partial_\mu\partial_\beta\bar{\omega}^{ab}_\gamma+\nonumber\\
    &+&\frac{1}{4}G^\alpha_{ab}\epsilon_{\alpha\nu}^{\phantom{\alpha\nu}\sigma\beta}z^{ab\gamma}_{\beta c}\partial_\sigma h^c_\gamma+\nonumber\\
    &+&\frac{1}{2k}\eta^\mu_a\eta^\alpha_b\epsilon_{\alpha\nu}^{\phantom{\alpha\nu}\sigma\beta}z^{ab\gamma}_{\beta c}\partial_\mu\partial_\sigma h^c_\gamma\;.\label{master1lin2}
\end{eqnarray}

We proceed with the gauge invariant fermionic Bargmann-Wigner equation \eqref{master2b}. Starting with the current itself, namely the second of \eqref{sc2}, its weak field approximation reads
\begin{eqnarray}
    F^{\mu\nu}&\approx&\frac{1}{2}\bar{\psi}\gamma^\mu\gamma^5\sigma^{\nu\alpha}\left(\bar{D}_\alpha-\frac{i}{4}\sigma_{mn}z^{mn\sigma}_{\alpha p}h^p_\sigma\right)\psi+\nonumber\\
    &-&\frac{1}{2}\left(\eta^\nu_ah_b^\alpha+\eta^\alpha_bh_a^\nu\right)\bar{\psi}\gamma^\mu\gamma^5\sigma^{ab}\partial_\alpha\psi+\nonumber\\
    &-&\frac{1}{2}h^\mu_c\bar{\psi}\gamma^c\gamma^5\sigma^{\nu\alpha}\partial_\alpha\psi+\frac{1}{2}\mathrm{h}\bar{\psi}\gamma^\mu\gamma^5\sigma^{\nu\alpha}\partial_\alpha\psi\;,\nonumber\\
    \label{bbw2}
\end{eqnarray}
with $\gamma^\mu=\eta^\mu_a\gamma^a$. We also need the linear approximation of the projected torsion defined in  \eqref{tensors1},
\begin{eqnarray}
     T^\mu_{ab}&\approx&-\eta^\nu_a\partial_\nu h^\mu_b+\eta^\nu_b\partial_\nu h^\mu_a-\eta^\nu_a\eta^\mu_c\bar{\omega}_{b\phantom{c}\nu}^{\phantom{b}c}+\nonumber\\
     &+&\eta^\nu_b\eta^\mu_c\bar{\omega}_{a\phantom{c}\nu}^{\phantom{a}c}
     -\eta^\nu_a\eta^\mu_cz_{b\nu d}^{\phantom{b}c\sigma}h_\sigma^d+\eta^\nu_b\eta^\mu_cz_{a\nu d}^{\phantom{a}c\sigma}h_\sigma^d\;.\nonumber\\
     \label{toraional1f1}
\end{eqnarray}
We still need to work on the sources appearing on the right hand side of equation \eqref{master2b}. They read
\begin{eqnarray}
    S_\mu^{ab}&\approx&\left(1+\mathrm{h}\right) \bar{S}^{ab}_\mu+h_\mu^c\bar{\psi}\gamma_c\gamma^5\sigma^{ab}\psi\;,\nonumber\\
    \Theta^{ab}_5&\approx&m\left(1+\mathrm{h}\right)\bar{\Theta}^{ab}_5\;,\nonumber\\
    X^\nu&\approx&\frac{i}{2}\left[\partial^\nu\left(-h^\mu_a+\mathrm{h}\eta^\mu_a\right)+\bar{\omega}_a^{\phantom{a}b\nu}\eta^\mu_b+\right.\nonumber\\
    &+&\left.z_{a\phantom{b}d}^{\phantom{a}b\nu\sigma}h^d_\sigma\eta^\mu_b\right]X^{a5}_\mu\;.\label{tensors2}
\end{eqnarray}
with
\begin{eqnarray}
    \bar{S}^{ab}_\mu&=&\bar{\psi}\gamma_\mu\gamma^5\sigma^{ab}\psi\;,\nonumber\\
    \bar{\Theta}^{ab}_5&=&\bar{\psi}\gamma^5\sigma^{ab}\psi\;,\nonumber\\
    X^{a5}_\mu&=&\bar{\psi}\gamma^5\gamma^a\partial_\mu\psi\;,
\end{eqnarray}
being independent of the gravitational fields. Therefore, the broken equation \eqref{master2b} assumes the form
\begin{eqnarray}
    \partial_\mu F^{\mu\nu}&=&\frac{i}{8}\left(\partial^\mu\bar{\omega}_{ab}^\nu-\partial^\nu\bar{\omega}_{ab}^\mu+z^{\sigma\nu}_{abc}\partial^\mu h^c_\sigma+\right.\nonumber\\
    &-&\left.z^{\sigma\mu}_{abc}\partial^\nu h^c_\sigma\right)\bar{S}^{ab}_\mu+\frac{im}{4}\left[\eta^\mu_a\partial_\mu h^\nu_b+\right.\nonumber\\
    &-&\left.\eta^\mu_b\partial_\mu h^\nu_a+\frac{i}{4}\left(\eta^\nu_a\eta^\mu_c\bar{\omega}_{b\phantom{c}\nu}^{\phantom{b}c}-\eta^\nu_b\eta^\mu_c\bar{\omega}_{a\phantom{c}\nu}^{\phantom{a}c}+\right.\right.\nonumber\\
    &+&\left.\left.\eta^\nu_a\eta^\mu_cz_{b\nu d}^{\phantom{b}c\sigma}h_\sigma^d-\eta^\nu_b\eta^\mu_cz_{a\nu d}^{\phantom{a}c\sigma}h_\sigma^d\right)\right]\bar{\Theta}^{ab}_5+\nonumber\\
    &+&\frac{i}{2}\left[\partial^\nu\left(-h^\mu_a+\mathrm{h}\eta^\mu_a\right)+\bar{\omega}_a^{\phantom{a}b\nu}\eta^\mu_b+\right.\nonumber\\
    &+&\left.z_{a\phantom{b}d}^{\phantom{a}b\nu\sigma}h^d_\sigma\eta^\mu_b\right]X^{a5}_\mu\;.\label{master2blin1}
\end{eqnarray} 

As mentioned before, equation \eqref{master2blin1} also works for external gravitational fields. This means that we can choose freely $\bar{\omega}$ and $h$. For instance, we can set a Riemannian weak background by setting $\bar{\omega}=0$. In this scenario one can study, for instance, the effect on spin-currents of gravitational waves or due to a Newtonian potential.

\section{Conclusions}\label{FINAL}

In this work we derived the master equations governing the interaction between gravitational fields and gauge invariant spin-currents as broken continuity equations given by \eqref{master1} and \eqref{master2b}. The corresponding equations in the weak field approximation are given by equations \eqref{master1lin2} and \eqref{master2blin1}. Our results are valid for weakly interacting charged fermions, in the electromagnetic sense. It is clear from our results that curvature plays an explicit influence on the non-conservation of the spin-currents. Torsion, however, only influences the fermionic sector. Moreover, since the derivative of the vierbein does not appear in the starting action, it does not contribute to the bosonic spin-current. Thence, the contribution for the bosonic spin-current comes exclusively from the spin-connection. This is a particularity of the Einstein-Hilbert action in the first order formalism. Another interesting point is that the gauge current couples to curvature in order to spoil the conservation of the bosonic gauge invariant spin-current. On the fermionic gauge invariant spin-current, it is the chiral current-curvature coupling that contributes to the non conservation of the current. This coupling is similar to the electromagnetic case \cite{Sobreiro:2011ie,Santos:2012ar} Furthermore, torsion couples to the current $\Theta$ defined in \eqref{tensors1}. In fact, the current $\Theta$, up to a normalization factor, appear in the right hand side of the non conservation equation of the second chiral current, see \eqref{chi4}. 

One more comment can be made, the inclusion of a cosmological constant term $\sim\Lambda\mathrm{e}$ will not affect any of our results because it does not contain any field derivatives (so it will not contribute to the Bargmann-Wigner current). Moreover, a cosmological constant term only affects the field equation of the vierbein, which is not used in the obtainment of the master equations. 

Our results remain at formal level in this work. Nevertheless, equations \eqref{master1} and \eqref{master2b} (and their linear versions  \eqref{master1lin2} and \eqref{master2blin1}) compose the essence for applications in order to find new physical effects on spin-currents under the influence of gravitational fields. An important observation is that these equations are valid for dynamical gravity systems as well as for non-dynamical gravitational fields (fixed background fields). These applications are left for future investigations. Obvious applications concern gravitational waves and black holes. But effects on Earth gravity, although probably weak, are not ruled out.

\section*{Acknowledgments}

This study was financed by The Coordenação de Aperfeiçoamento de Pessoal de Nível Superior - Brasil
(CAPES) - Finance Code 001.

\bibliographystyle{utphys2}
\bibliography{library}

\end{document}